\documentstyle[aps,prb,epsfig,twocolumn]{revtex}

\begin{document}

\bibliographystyle{prsty}

\draft

\wideabs{
\title{ Towards the Understanding of Stability Puzzles in Phage $\lambda$  }
\author{ P. Ao$^{1)}$ and L. Yin$^{2)}$  }
\address{ $^{1)}$ Departments of Physics and Mechanical Engineering, 
   University of Washington, Seattle, WA 98195        \\
   $^{2)}$Department of Physics, Peking University, Beijing, PR China   }
\address{ }
\date{ submitted to PRL on July 21 2003 }

\maketitle


}


In their elegant work\cite{aurell} Aurell and Sneppen developed a framework to quantify the robustness of a biological state.  They have successfully applied their formulation to the robustness of the epigenetic state of the wild type phage $\lambda$, but concluded, and supported by their further numerical study\cite{aurell2}, that their model is not robust against various mutations in the gene regulatory network of phage $\lambda$. They called this disagreement between their mathematical modeling and biological experiments the stability puzzle. We agree with their assessment that the robustness can be viewed as a first exit problem and do not dispute the possibility that certain unknown biology may be responsible for their discrepancy. In this comment we would like, however, to call the attention to two important aspects, one on biology and one on modeling, overlooked by Aurell and Sneppen, and to argue that the including of those two aspects may be sufficient to understand the stability puzzle.

{\bf Biological aspect}: {\it in vivo vs. in vitro}.  It is well known that so far most molecular parameters in biology are measured {\it in vitro}, that is, not in the native chemical and biological environments of the living organism ({\it in vivo}).  This is same for phage $\lambda$, one of most well studied organisms.\cite{phage} Tremendous efforts have been made to ensure the {\it in vitro} conditions as close as those of {\it in vivo}.  Subsequently, it has been generally concluded that though there is indeed an {\it in vivo} and {\it in vitro} difference, this difference is small and normally not exceeds 30$\%$ of the measured {\it in vitro} value.\cite{ptashne} Such a small difference, for the case of affinities in DNA-protein binding dominated the dynamics of gene regulatory networks, is about a few kcal/mol. Regarding to the usual qualitative nature of biological data where measurements for the same quantity are common to be differed by a factor of order unity, one would be satisfied to have such {\it in vitro} values. Unfortunately, for the stability as a first exit problem, the escape rate can be exponentially sensitive to some parameters. Hence 10$\%$ change in a critical parameter can result in orders of magnitudes change in the escape rate.  In physics and chemistry such examples are too abundant to enumerate here. This comes our first observation to understand the stability of the biological states: 
One must carefully take the in vivo and in vitro difference into account for dominant molecular parameters in the modeling. This has been attempted but not taken seriously in Ref.[1] and [2].

{\bf Modeling aspect}: the nature of noise.   As it is clear from the mathematical modeling, the noise strength equivalent to temperature plays the most important role, as well exemplified by the classical Arrhenius law. We found two important features on noise which have not be taken cared of properly in Ref.[1], the diffusion approximation for the intrinsic noise and the presence of extrinsic noise.  First, Aurell and Sneppen modeled the noise as it directly comes from the chemical reaction rates. Indeed, the chemical reactions are noisy processes. What used by Aurell and Sneppen is the standard diffusion approximation to obtain the noise strength, called the intrinsic noise.\cite{vankampen} This approximation is good near a stable fixed point, but can be very problematic when applied to relative stability problem. Orders of magnitudes in error may be embedded in this diffusion approximation when calculating the escape rate.\cite{gaveau}  More precise approximation is needed to get a better description of the intrinsic noise,\cite{hanggi} which we believe can be done with an improved diffusion approximation.  The second feature is more seriously, regarding to the extrinsic noise: There are other noisy processes beyond those dictated by reaction rates of the deterministic part of the modeling. In fact, biological experimental data already indicated the existence of the extrinsic noise.\cite{elowitz}  The extrinsic noise is not in the modeling of Aurell and Sneppen, which may well account for their inability to solve the stability puzzle.

With the consideration of above two biological and modeling aspects, the {\it in vivo} and {\it in vitro} difference and the proper modeling of the noise, can one understand the stability puzzle in phage $\lambda$?  It is evident that those two aspects will give enough difference in the numerical outcome of the modeling. One should not be surprised to find the positive answer.

\end{document}